\documentstyle[prb,aps,epsfig,floats]{revtex}

\begin{document}

\twocolumn
\psfull
\draft

\title{Quantum transport in ballistic conductors: evolution from conductance
quantization to resonant tunneling}
\author{Branislav K. Nikoli\' c and Philip B. Allen}
\address{Department of Physics and Astronomy,
SUNY at Stony Brook, Stony Brook, New York 11794-3800}

\maketitle

\begin{abstract}
We study the transport properties of an atomic-scale contact in
the ballistic regime. The results for the conductance and related
transmission eigenvalues show how the properties of the ideal
semi-infinite leads (i.e. measuring device) as well as the
coupling between the leads and the conductor influence the
transport in a two-probe geometry. We observe the evolution from
conductance quantization to resonant tunneling conductance peaks
upon changing the hopping parameter in the disorder-free
tight-binding Hamiltonian which describes the leads and the
coupling to the sample.
\end{abstract}


Mesoscopic physics~\cite{datta} has changed  our understanding of
transport in condensed matter systems. The discovery of new
effects, such as weak localization~\cite{wl} or universal
conductance fluctuations,~\cite{ucf} has been accompanied by
rethinking of the established ideas in a new light. One of the
most spectacular discoveries of mesoscopics is conductance
quantization (CQ)~\cite{wees,wharam} in a short and narrow
constriction connecting two high-mobility (ballistic)
two-dimensional electron gases. The conductance of these quantum
point contacts as a function of the constriction width $W \sim
\lambda_F$ has steps of magnitude $2e^2/h$. New experimental
techniques have allowed observation of similar
phenomena~\cite{ruitenbeek} in metallic point contacts of atomic
size. The Landauer formula~\cite{landauer} for the two-probe
conductance
\begin{equation}\label{eq:landauer}
  G=\frac{2e^2}{h} \, \text{Tr} \, ({\bf t t}^{\dag})
  = G_Q \sum_{n=1}^M T_n,
\end{equation}
has provided an explanation of the stepwise conductance in terms
of the number $N \leq M$ of transverse propagating states
(``channels'') at the Fermi energy $E_F$ which are populated in
the constriction. Here ${\bf t}$ is the transmission matrix,
$T_n$ transmission eigenvalues and $G_Q=2e^2/h$ is the
conductance quantum. In the ballistic case $({\bf t
t}^{\dag})_{ij}$ is $\delta_{ij}$, or equivalently $T_n$ is $1$.
Further studies have explored CQ under a range of
conditions.~\cite{sarma} They include
geometry,~\cite{glazman,stone} scattering on
impurities,~\cite{maslov} temperature effects, and magnetic field.

In this paper we study the influence of the attached leads on
ballistic transport ($\ell > L$, $\ell$ being elastic mean free
path, $L$ being the system size) in a nanocrystal. We assume that
in the two-probe theory an electron leaving the sample does not
reenter the sample in a phase-coherent way. This means that at
zero temperature phase coherence length $L_\phi$ is equal to the
length of the sample $L$. In the jargon of quantum measurement
theory, the leads act as a ``macroscopic measurement apparatus''.
Our concern with the influence of the leads on conductance is
therefore also a concern of quantum measurement theory. Recently,
the effects of a lead-sample contact on quantum transport in
molecular devices have received increased attention in the
developing field of ``nanoelectronics''.~\cite{ventra} Also, the
simplest lattice model and related real-space Green function
technique are chosen here in order to address some practical
issues which appear in the frequent use of these
methods~\cite{datta} to study transport in disordered samples. We
emphasize that the relevant formulas for transport coefficients
contain three different energy scales (corresponding to the lead,
the sample, and the lead-sample contact), as discussed below.

\begin{figure}
\centerline{
\psfig{file=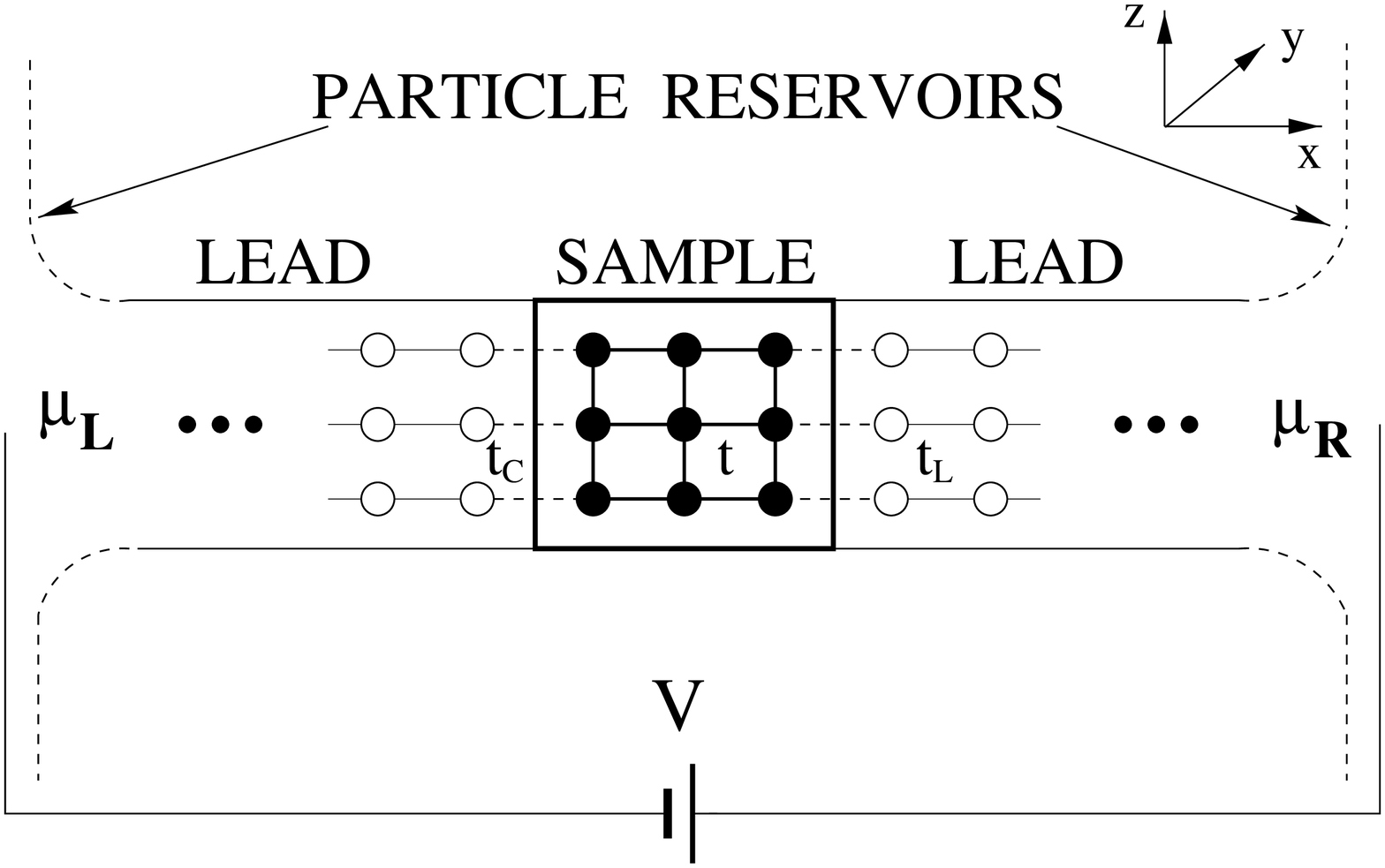,height=2.3in,width=3.0in,angle=0} }
\vspace{0.2in} \caption{A two dimensional version of our actual
3D model of a two-probe measuring geometry. Each site hosts a
single s-orbital which hops to six (or fewer for surface atoms)
nearest neighbors. The hopping matrix element is $t$ (within the
sample), $t_{\text L}$ (within the leads), and $t_{\text C}$
(coupling of the sample to the leads). The leads are
semi-infinite and connected at $\pm \infty$ to reservoirs with
potential difference $\mu_L-\mu_R=eV$.} \label{fig:setup}
\end{figure}

In order to isolate only these effects we pick the strip geometry
in the two-probe measuring setup shown on Fig.~\ref{fig:setup}.
The nanocrystal  (``sample'') is placed between two ideal (disorder-free)
semi-infinite ``leads'' which are connected to macroscopic reservoirs.
The electrochemical potential
difference $e V=\mu_L-\mu_R$ is measured between the reservoirs. The leads have
the same cross section as the sample. This eliminates scattering induced by the
wide to narrow geometry~\cite{stone} of the sample-lead interface. The whole
system is described by a clean tight-binding Hamiltonian (TBH) with nearest
neighbor hopping parameters $t_{\bf mn}$
\begin{equation}\label{eq:tbh}
  \hat{H} = \sum_{\langle {\bf m},{\bf n} \rangle} t_{{\bf m} {\bf n}}
  |{\bf m} \rangle \langle {\bf n}|,
\end{equation}
where $| {\bf m} \rangle$ is the orbital $\psi({\bf r}-{\bf m})$
on the site ${\bf m}$. The ``sample'' is the central section with
$N_x \times N_y \times N_z$ sites. The ``sample'' is perfectly
ordered with $t_{\bf mn}=t$. The leads are the same except
$t_{\bf mn}=t_{\text L}$. Finally, the hopping parameter
(coupling) between the sample and the lead is $t_{\bf
mn}=t_{\text C}$. We use hard wall boundary conditions in the
$\hat{y}$ and $\hat{z}$ directions. The different hopping
parameters introduced here have to be used to get the conductance
at Fermi energies throughout the whole band extended by the
disorder,~\cite{nikolic1} i.e. $t_{\text L} > t$. Thus, one has to
be aware of the conductances we calculate in our analysis when
engaging in such studies.

\begin{figure}
\centerline{ \psfig{file=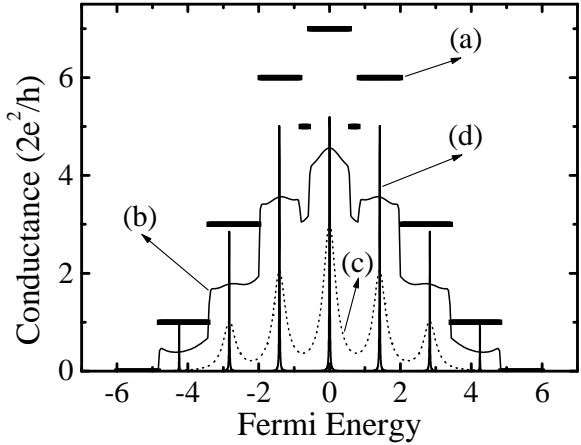,height=3.0in,angle=-90} } \vspace{0.2in}
\caption{Conductance G of an atomic-scale ballistic contact $3 \times 3 \times 3$
for the following values of lead and coupling parameters: (a) $t_{\text C}=1,
t_{\text L}=1$, (b) $t_{\text C}=1.5$, $t_{\text L}=1$ (c) $t_{\text C}=3$,
$t_{\text L}=1$, and (d) $t_{\text C}=0.1, t_{\text C}=1$. In the case (d) the
conductance peaks are connected by smooth curves of $G < 0.004 e^2/h$.}
\label{fig:tunnel}
\end{figure}

\begin{figure}
\centerline{ \psfig{file=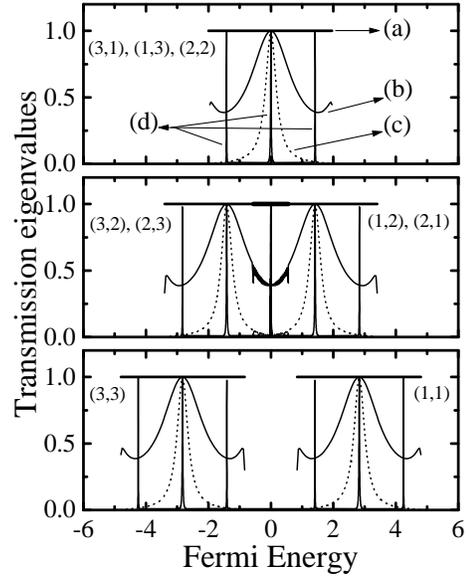,height=3.0in,angle=0} } \vspace{0.2in}
\caption{Transmission eigenvalues of the atomic-scale ballistic contact $3 \times
3 \times 3$. The parameters $t_{\text L}$ and $t_{\text C}$ are the same as in
Fig.~\ref{fig:tunnel}. All channels $(i,j) \equiv (k_y(i),k_z(j))$ whose subbands
are identical have the same $T_n$. This gives the degeneracy of $T_n$: three
(upper panel), two (middle panel), and one (bottom panel). In the middle panel
the lower two subbands have an energy interval of overlap with the upper two
subbands.} \label{fig:trans}
\end{figure}

Our toy model shows exact conductance steps in multiples of $G_Q$
when $t_{\text C}=t_{\text L}=t$. This is a consequence of
infinitely smooth (``ideally adiabatic''~\cite{glazman})
sample-lead geometry. Then we study the evolution of quantized
conductance into resonant tunneling conductance while changing
the parameter $t_{\text L}$ of the leads as well as the coupling
between the leads and the conductor $t_{\text C}$. An example of
this evolution is given on Fig.~\ref{fig:tunnel}. The equivalent
evolution of the transmission eigenvalues $T_n$ of channels is
shown on Fig.~\ref{fig:trans}. A similar evolution has been
studied recently in one-atom point contacts.~\cite{yamaguchi}

The non-zero resistance is a purely geometrical effect~\cite{imry}
caused by reflection when the large number of channels in the
macroscopic reservoirs matches the small number of channels in
the lead. The sequence of steps ($1,3,6,5,7,5,6,3,1$ multiples of
$G_Q$ as the Fermi energy $E_F$ is varied) is explained as
follows. The eigenstates in the leads, which comprise the
scattering basis,  have the form $\psi_{\bf k} \propto \sin (k_y
m_y) \sin (k_z m_z) e^{i k_x m_x}$ at atom ${\bf m}$, with energy
$E=2t_{\text L}[\cos (k_x a)+\cos (k_y a) + \cos (k_z a)]$, where
$a$ is the lattice constant. The discrete values $k_y(i)=i
\pi/(N_y+1)a$ and $k_z(j)=j \pi/(N_z+1)a$  define subbands or
``channels'' labeled by $(k_y,k_z) \equiv (i,j)$, where $i$ runs
from $1$ to $N_y$ and $j$ runs from $1$ to $N_z$. The channel
$(k_y,k_z)$ is open if $E_F$ lies between the bottom of the
subband, $2t_{\text L}[-1+\cos (k_y a) + \cos (k_z a)]$, and the
top of the subband, $2 t_{\text L} [1+\cos (k_y a) + \cos (k_z
a)]$. Because of the degeneracy of different transverse modes in
3D, several channels $(k_y,k_z)$ open or close at the same
energy. Each channel contributes one conductance quantum $G_Q$.
This is shown on Fig.~\ref{fig:tunnel} for a sample with $3
\times 3$ cross section where the number of transverse propagating
modes is $M=9$. In the adiabatic geometry, channels do not mix,
i.e. the transmission matrix is diagonal in the basis of channels
defined by the leads.

We compute the conductance using the expression obtained in the
framework of Keldysh technique by treating the coupling between
the central region and the lead as a perturbation.~\cite{caroli}
This provides the following, Landauer-type, formula for the
conductance in the non-interacting system
\begin{eqnarray}\label{eq:greenlandauer}
  G & = & \frac{2e^2}{h} \, \text{Tr} \left (\hat{\Gamma}_L \, \hat{G}^{r}_{1 N_x} \,
  \hat{\Gamma}_R \, \hat{G}^{a}_{N_x 1} \right ) = \frac{2e^2}{h} \, \text{Tr} \, ({\bf t t}^{\dag}), \\
  {\bf t} & = & \sqrt{\hat{\Gamma}_L} \, \hat{G}^{r}_{1 N_x} \sqrt{\hat{\Gamma}_R}
  \label{eq:ttgreen}.
\end{eqnarray}
Here $\hat{G}^{r}_{1 N_x}$, $\hat{G}^{a}_{N_x 1}$ are matrices
whose elements are the Green functions connecting the layer $1$
and $N_x$ of the sample. Thus only the block $N_y \times N_z$ of
the whole matrix $\hat{G}({\bf n},{\bf m})$ is needed to compute
the conductance. The positive operator
$\hat{\Gamma}_L=i(\hat{\Sigma}_L^r-\hat{\Sigma}_L^a)=-2\,
\text{Im} \, \hat{\Sigma_L} > 0$ is the counterpart of the
spectral function $\hat{A}=i(\hat{G}^r - \hat{G}^a)$ for the
self-energy $\hat{\Sigma}_L$ introduced by the left lead. It
``measures'' the coupling of the open sample to the left lead
($\hat{\Gamma}_R$ is equivalent for the right lead).The Green
operator is defined as the inverse of $(E-\hat{H})$ including the
relevant boundary conditions. Instead of inverting the infinite
matrix we invert only $(E-\hat{H}_S)$ defined on the Hilbert
space spanned by orbitals $|{\bf m} \rangle$ inside the
sample~\cite{caroli}
\begin{equation}\label{eq:green}
  \hat{G}^{r}=(E-\hat{H}_S - \hat{\Sigma}^{r})^{-1},
\end{equation}
where $\hat{H}_S$ is TBH for the sample only. This is achieved by
using the retarded self energy
$\hat{\Sigma}^{r}=\hat{\Sigma}_L^{r}+\hat{\Sigma}_R^{r}$
introduced by the left $(L)$ and the right $(R)$ lead. In site
representation Green operator $\hat{G}^{r,a}$ is a Green function
matrix $\hat{G}^{r,a}({\bf n},{\bf m})=\langle {\bf n} |
\hat{G}^{r,a} | {\bf m} \rangle$. Equation~(\ref{eq:green}) does
not need the small imaginary part $i 0^+$ necessary to specify
the boundary conditions for the retarded or advanced Green
operator $\hat{G}^{r,a}$ because the lead self-energy
$(\hat{\Sigma}^{a}=[\hat{\Sigma^{r}}]^{\dagger})$ adds a well
defined imaginary part to $E-\hat{H}_S$. This imaginary part is
related to the average time an electron spends inside the sample
before escaping into the leads. The self-energy terms have
non-zero matrix elements only on the edge layers of the sample
adjacent to the leads. They are given~\cite{datta} in terms of
the Green function on the lead edge layer and the coupling
parameter $t_{\text C}$
\begin{eqnarray}\label{eq:sigmann}
  \hat{\Sigma}^{r}_{L,R}({\bf n},{\bf m}) & = & \frac{2}{N_y+1} \frac{2}{N_z+1} \sum_{k_y,k_z}
  \sin( k_y n_y ) \sin(k_z n_z) \nonumber \\
  & & \times \hat{\Sigma}^r (k_y,k_z) \sin( k_y m_y ) \sin(k_z
  m_z),
\end{eqnarray}
where  $({\bf n},{\bf m})$ is the pair of sites on the surfaces
inside the sample which are adjacent to the leads ($L$ or $R$).
The self-energy $\hat{\Sigma}^r (k_y,k_z)$ in the channel
$(k_z,k_y)$ is given by
\begin{equation}\label{eq:sigmakk1}
  \hat{\Sigma}^{r}(k_y,k_z)=\frac{t_{\text C}^2}{2t_{\text L}^2} \left (E_{\Sigma}-i
  \sqrt{4t_{\text L}^2-E_{\Sigma}^2} \right ),
\end{equation}
for $|E_{\Sigma}|<2t_{\text L}$. We use the shorthand notation
$E_\Sigma=E-\varepsilon(k_y,k_z)$, where $\varepsilon(k_y,k_z)=2t_{\text L} [
\cos(k_y a) + \cos(k_z a) ]$ is the energy of quantized transverse levels in the
lead. In the opposite case $|E_{\Sigma}|>2t_{\text L}$ we have
\begin{equation}\label{eq:sigmakk2}
  \hat{\Sigma}^{r}(k_y,k_z)=\frac{t_{\text C}^2}{2t_{\text L}^2} \left ( E_{\Sigma}-\text{sgn}
  \, E_{\Sigma} \sqrt{E_{\Sigma}^2-4t_{\text L}^2} \right ).
\end{equation}

In order to study the conductance as a function of two parameters $t_{\text L}$
\begin{figure}
\centerline{ \psfig{file=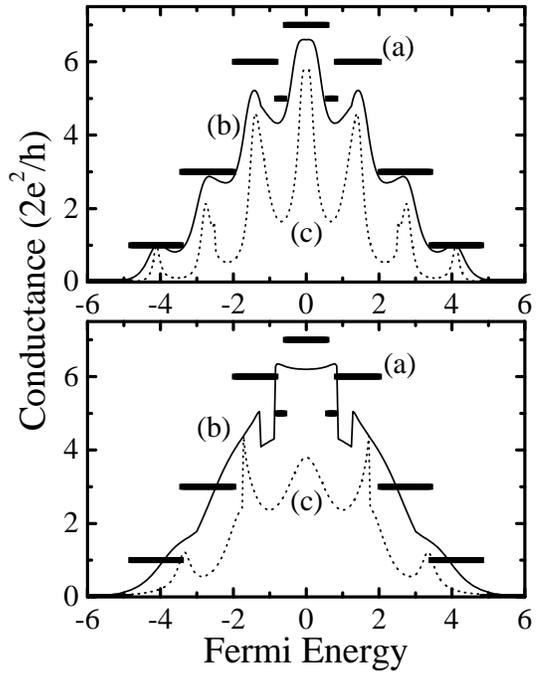,height=3.5in,angle=0} } \vspace{0.2in}
\caption{Conductance $G$ of the atomic-scale ballistic conductor $3 \times 3 \times
3$ for the following values of lead and coupling parameters: Upper panel --- (a)
$t_{\text C}=1$, $t_{\text L}=1$, (b) $t_{\text C}=1$, $t_{\text L}=1.5$, and (c)
$t_{\text C}=1$, $t_{\text L}=3$; Lower panel --- (a) $t_{\text C}=1$, $t_{\text
L}=1$, (b) $t_{\text C}=1.5$, $t_{\text L}=1.5$, and (c) $t_{\text C}=3$, $t_{\text
L}=3$. } \label{fig:plain}
\end{figure}
and $t_{\text C}$ we change either one of them while holding the
other fixed (at the unit of energy specified by $t$), or both at
the same time. The first case is shown on Fig.~\ref{fig:tunnel}
and Fig.~\ref{fig:plain} (upper panel),  while the second one on
Fig.~\ref{fig:plain} (lower panel). The conductance is depressed
in all cases since these configurations of hopping parameters
$t_{\bf mn}$ effectively act as a barriers. There is a reflection
at the sample-lead interface due to the mismatch of the subbands
in the lead and in the sample when $t_{\text L}$ differs from
$t$. This demonstrates that adiabaticity is not necessary
condition for CQ. In the general case, each set of channels which
have the same energy subband is characterized by its own
transmission function $T_n(E_F)$. When the coupling $t_{\text
C}=0.1$ is small a double-barrier structure is obtained which has
a resonant tunneling conductance. The electron tunnels from one
lead to the other via discrete eigenstates. The transmission
function is composed of peaks centered at $E_r=2t[\cos (k_x
a)+\cos (k_y a) + \cos (k_z a)]$, where $k_x=k \pi/(N_x+1)a$ is
now quantized inside the sample, i.e. $k$ runs from $1$ to $N_x$.
The magnitude and width of peaks is defined by the rate at which
an electron placed between barriers leaks out into the lead.
These rates are related to the level widths generated through the
coupling to the leads. In our model they are energy (i.e. mode)
dependent. For example at $E_F=0$ seven transmission eigenvalues
are non-zero (in accordance with open channels on
Fig.~\ref{fig:trans}) and exactly at $E_F=0$ three of them have
$T=1$ and four $T=0.5$. Upon decreasing $t_{\text C}$ further all
conductance peaks, except the one at $E_F=0$, become negligible.
Singular behavior of $G(E_F)$ at subband edges of the leads was
observed before.~\cite{maslov}

It is worth mentioning that the same results are obtained using a non-standard
version of Kubo-Greenwood formula~\cite{kubo} for the volume averaged conductance
\begin{mathletters}
\label{eq:kubo}
\begin{eqnarray}
 G & = & \frac{4e^2}{h} \frac{1}{L_x^2} \, \text{Tr} \left (\hbar \hat{v}_x
 \text{Im} \,  \hat{G} \, \hbar \hat{v}_x \text{Im} \, \hat{G} \right ), \\
 \text {Im} \, \hat{G} & = & \frac{1}{2i} (\hat{G}^r - \hat{G}^a),
\end{eqnarray}
\end{mathletters}
where $v_x$ is the $x$ component of the velocity operator. This was originally
derived for an infinite system without any notion of leads and reservoirs. The
crucial non-standard aspect is use of the Green function~(\ref{eq:green}) in
formula~(\ref{eq:kubo}). This takes into account, through lead self-energy~(\ref{eq:sigmann}), the boundary conditions at the reservoirs. The reservoirs are necessary in both Landauer
and Kubo formulations of linear transport for open finite systems. They provide
thermalization and thereby steady state of the transport in the central region.
Semi-infinite leads~\cite{lee} are a convenient method to model the macroscopic reservoirs.
When employing the Kubo formula~(\ref{eq:kubo}) one can use current conservation
and compute the trace only on two adjacent layers inside the sample. To get the
correct results in this scheme $L_x$ in Eq.~(\ref{eq:kubo}) should be
replaced~\cite{nikolic1} by a lattice constant $a$.

In the quantum transport theory of disordered systems the
influence of the leads on the conductance of the sample is
understood as follows.~\cite{hans} An isolated sample has a
discrete energy spectrum. Attaching leads necessary for transport
measurements will broaden energy levels. If the level width
$\Gamma$ due to the coupling to leads is larger than the Thouless
energy $E_{\text{Th}}=\hbar/\tau_{\text{D}} \simeq \hbar D/L^2$,
($D=v_F \ell/3$ being the diffusion constant) the level
discreteness is unimportant for transport. For our case of
ballistic conduction, $E_{\text{Th}}$ is replaced by the inverse
time of flight $\hbar v_F/L$. In the disordered sample where
$\Gamma \gg E_{\text{Th}}$, varying the strength of the coupling
to the leads will not change the transport coefficients. In other
words, the intrinsic resistance of the sample is much larger than
the resistance of the lead-sample contact.~\cite{efetov} In the
opposite case, discreteness of levels becomes important and the
strength of the coupling defines the conductance. This is the
realm of quantum dots~\cite{carlo} where weak enough coupling can
make the charging energy $e^2/2C$ of a single electron important
as well. Changing the properties of the dot-lead contact affects
the conductance, i.e. the result of measurement depends on the
measuring process. The decay width
$\Gamma=\hbar/\tau_{\text{dwell}}$ of the electron emission into
one of the leads is determined by transmission probabilities of
channels through the contact and mean level spacing.~\cite{hans}
This means that mean dwell time $\tau_{\text{dwell}}$ inside our
sample depends on both $t_{\text C}$ and $t_{\text L}$. Changing
the hopping parameters will make $\tau_{\text{dwell}}$ greater
than the time of flight $\tau_f=L/v_F$. Thus we find that
ballistic conductance sensitively depends on the parameters of
the dephasing environment (i.e. the leads).

In conclusion, we have studied the transport properties of a
ballistic nanocrystal placed between two semi-infinite leads in
the simplest strip geometry. We observe extreme sensitivity of
the conductance to changes in the hopping parameter in the leads
as well as the coupling between the leads and the sample. As can
be easily anticipated, the conductance evolves from perfect
quantization (as a result of an ideal adiabatic geometry) to
resonant tunneling. Nevertheless, it is quite amusing that vastly
different $G(E_F)$ are obtained between these two limits (e.g.
Fig.~\ref{fig:plain}). The results are of relevance for the
analogous theoretical studies in disordered conductors as well as
in the experiments using clean metal junctions with different
effective electron mass throughout the circuit.

This work was supported in part by NSF grant no. DMR 9725037. We
thank I. L. Aleiner for interesting discussions.

\end{document}